# Cavity-enhanced acousto-optic modulators on polymer-loaded lithium niobate integrated platform

*Zhi Jiang,[1] Danyang Yao,[1*] Xu Ran,[1] Yu Gao,[1] Jianguo Wang,[2] Xuetao Gan,[2*] Yan Liu,[1,3] Yue Hao,[1] and Genquan Han[1,3]*

[1] State Key Laboratory of Wide-Bandgap Semiconductor Devices and Integrated Technology, School of Microelectronics, Faculty of integrated circuits, Xidian University, Xi'an, 710071, China;

[2] Key Laboratory of Light Field Manipulation and Information Acquisition, Ministry of Industry and Information Technology, and Shaanxi Key Laboratory of Optical Information Technology, School of Physical Science and Technology, Northwestern Polytechnical University, Xi'an, 710129, China;

[3] Hangzhou Institute of Technology, Xidian University, Hangzhou, 311200, China;

**ABSTRACT:**

On chip acousto-optic (AO) modulation represents a significant advancement in the development of highly integrated information processing systems. However, conventional photonic devices face substantial challenges in achieving efficient conversion due to the limited overlap between acoustic waves and optical waves. In this study, we address this limitation by demonstrating an enhanced conversion effect of photonic crystal nanobeam cavities (PCNBCs) in AO modulation on a polymer-loaded lithium niobate integrated platform. Attributed to the high ratio of quality factor (Q) to mode volume (V) and optimal light-sound overlap within the nanocavity, PCNBCs-based AO modulator exhibits a significantly enhanced extinction ratio of 38 dB with a threshold RF power below -50 dBm, which is two orders of magnitude lower than that based on micro-ring resonator (MRRs). In addition, robust digital amplitude shift keying modulations using selected RF and optical channels of the PCNBCs-enhanced AO modulators. These findings validate the compelling properties of the PCNBCs photonic platform, establishing it as a promising candidate for on-chip integrated microwave photonics, optical transceivers, and computing applications.

# ■ INTRODUCTION

Interactions between localized acoustic and optical waves on a chip provide a highly effective method for converting electrical signals into optical signals, opening the pathway towards acousto-optic (AO) devices, including modulators[1-4], frequency shifters[5,6], non-reciprocal components[7-9], matrix-vector multiplications[10], and quantum information processors[11]. This delicate manipulation of photons involves controlling the refractive index of optical medium through elastic vibrations generated by interdigital transducers (IDTs). Recent advancements in acousto-optic (AO) modulators, such as suspended lamb-wave IDTs[12,13], superior elasto-optical material integrations[3,14], and sideband-resolved resonance coupling designs[15,16], are committed to improve the microwave-to-optical conversion efficiency. However, conventional photonic devices like Mach-Zehnder interferometers (MZIs) and micro-ring resonators (MRRs) face substantial challenges in achieving high-efficient conversion due to the insufficient field overlap between propagating acoustic waves and confined optical waves[17].

In contrast, photonic crystal nanobeam cavities (PCNBCs) offer a unique optical cavity structure capable of trapping photons in small volumes, thereby exhibiting strong modulation capabilities. These capabilities have been demonstrated in electro-optic (EO)[18,19], magnetooptic[20], carrier injection[21] and thermo-optic modulation[22,23]. Consequently, PCNBCs are particularly competitive in AO modulator studies for the following reasons: (i) PCNBCs are known for their exceptional quality factor to modal volume ratios (Q/V)[24,25], making them highly responsive to strain change generated by mechanical moving. (ii) The concentrated Gaussian-like field distribution[26] of the optical mode together with the straight waveguide profile of PCNBCs make it possible for the IDTs to cover the entire optical cavity region, enabling an overlap between acoustic waves and optical waves close to 100%. Despite these theoretical advantages, the enhanced AO modulation effects of PCNBCs have not yet been verified in experiments. To fully harness the benefits of PCNBCs in practical applications, further experimental validation is necessary to overcome performance bottlenecks in various existing AO modulators.

In this work, we present cavity-enhanced AO modulators developed on the lithium niobate (LN) integrated platform. Instead of directly etching the LN, we design and patterned waveguides and optical cavities to confine light modes by loading polymer dielectrics onto a planar LN-on-insulator (LNOI) substrate. Both high-Q MRR and PCNBC, operating in transverse electric (TE) polarization modes, are integrated with the same IDTs. Our theoretical calculations and experimental results demonstrate that PCNBC-based AO modulator exhibit significant advantages in terms of threshold RF power, conversion efficiency and signal-to-noise (SNR) ratios. These findings suggest that PCNBCs are a highly competitive technological tool for further improving the performance of AO modulator. Additionally, these cavities offer a promise of drastically reducing the RF power consumption and footprint of integrated devices, making them a promising signal modulation candidate for on-chip integrated microwave photonics, optical transceiver and computing applications.

■ **ACOUSTO-OPTIC CAVITY DESIGNS**

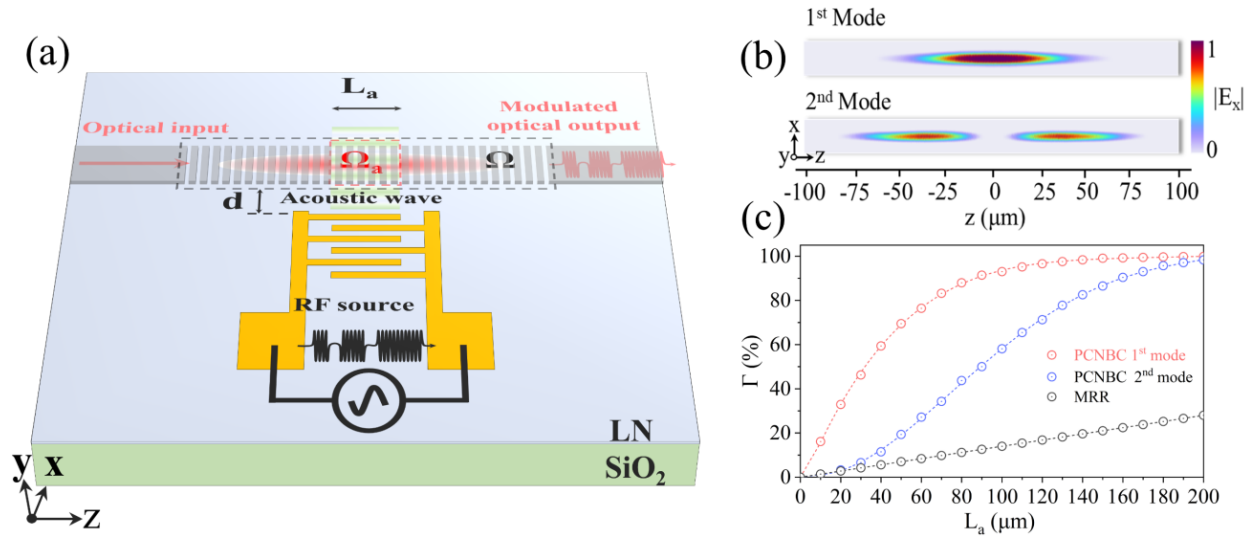

**Figure 1.** (a) Schematic illustration of the cavity-enhanced AO modulator. $\Omega_a$ is the AO interaction region. The acoustic wave is launched by IDT, with an aperture of $L_a$. The distance between the PCNBC and IDT is denoted by d. (b) The Gaussian-shaped electric field distribution of the PCNBC for the fundamental mode and the second-order mode, respectively. (c) Calculated confinement factors $\Gamma$ for PCNBC and MRR versus various aperture $L_a$.

LN stands out among piezoelectric material platforms like AlN, GaAs, and PZT due to its exceptionally high piezoelectric coefficient, low propagation loss[27,28]. These properties make LN to be a promising platform for AO coupling interactions. However, monolithic LN optical devices still suffer challenges in the high-quality etching process. The polymer-loaded LN integrated platform allows for the creation of an etchless rib-waveguide, capable of exciting phenomena such as low loss propagation[29], second harmonic generation (SHG)[30], mode multiplexer[31], as well as the implementation of components such as arrayed waveguide gratings (AWG)[32], EO modulators[19], and AO modulators[33,34]. Inspired by this, the etchless LN AO modulator consists of a one-dimensional PCNBC driven by an adjacent IDT for microwave-to-optical conversion is proposed as shown in Figure 1a. This device generates an acoustic wave along the x-axis orientation on a y-cut LNOI substrate, consisting of a 300 nm thick piezoelectric LN layer and a 2 μm SiO$_2$ layer. Then, all the patterned optical components are formed in the 400 nm polymer loading material on the LNOI substrate. Herein, $L_a$ represents the IDT emitting aperture, while $\Omega_a$ and $\Omega$ denote the AO interaction and the nanocavity regions, respectively. The piezoelectric properties of LN enable the generation of surface acoustic wave (SAW) when the applied RF signal matches the resonance frequency of the IDT. For a given $L_a$, it is desirable that the majority of photons are concentrated within the $\Omega_a$ region to enhance the interaction between the acoustic wave and optical wave. The proportion could be quantified as the confinement factor $\Gamma$, which is the ratio of the optical energy confined in $\Omega_a$ to that in the entire optical cavity $\Omega$, and expressed as,

$$\Gamma = \frac{\iiint_{\Omega_a} \frac{1}{2}\varepsilon_r(r)|E(r)|^2 dr}{\iiint_{\Omega} \frac{1}{2}\varepsilon_r(r)|E(r)|^2 dr} \tag{1}$$

where $\varepsilon_r(r)$ is the real part of the dielectric constant, and $E(r)$ is the electric field amplitude at spatial position r. For MRR, considering the uniform distribution of $E(r)$ along the ring waveguide, the expression of $\Gamma$ can be simplified to $\Gamma=2L_a/L_c$, where $L_c$ is the circumference of the ring cavity.

To verify our design, a 3D finite element method (FEM) simulation is performed using COMSOL Multiphysics. Figure 1b shows the simulation results of the optical electric field profiles of the PCNBC in both the fundamental mode and the second-order mode. The lattice constant and width of the designed PCNBC are 430 nm and 2 μm, respectively. Subsequently, the PCNBC is formed by a series of polymer dielectric blocks that defined by filling factors. According to filling factor, the PCNBC could be divided into taper region and mirror region. In the taper region, the filling factor varies parabolically from 0.43 at the center of PCNBC across 120 unit cells to 0.3 at the end. The mirror region consists of 130 unit cells that defined by the same filling factor of 0.3. (For detailed design information in high-Q PCNBC could be found in our previous work[35]). The fundamental mode, with Q factor of $5\times10^4$ and resonance wavelength at 1550 nm, exhibits a well-confined Gaussian-shaped electric field distribution concentrated in the center of the cavity. In contrast, the second-order mode, with Q factor of $0.6\times10^4$ and resonance wavelength at 1553 nm, degenerates into a two-lobed electric field distribution. The calculated $\Gamma$ as a function of $L_a$ for the fundamental mode, the second-order mode, and MRR is depicted in Figure 1c. One-dimensional PCNBC, which ensure a high-quality factor and small mode volume, can effectively couple with the acoustic waves emitted by IDT. For the fundamental mode, $\Gamma$ increases rapidly and approaches saturation when $L_a$ exceeds 100 μm. The second-order mode, characterized by a hollow-field pattern with two lobes, exhibits a slower increase in $\Gamma$ as the aperture widens. However, once the aperture begins to cover the two regions where energy is concentrated, $\Gamma$ experiences a swift increase and then proceeds to a saturation point. The distinct circular configuration of MRR poses challenges for achieving comprehensive optical cavity coverage by the acoustic emission aperture. In Figure 1c, for the MRR with a bending radius of 180 μm and a straight segment of 200 μm, $\Gamma$ exhibits a linear growth with the aperture size, as expected. Note that the maximum value of $\Gamma$ remains at only around 28%, even with an aperture as wide as 200 μm.

Within $\Omega_a$, the propagating SAW deforms the optical waveguide and generates a dynamic strain field. In this regime, AO interaction can be characterized by the single-photon coupling strength $g_0$. According to ref[16,36], we can derive $g_0$ as follow,

$$g_0 = \Gamma \omega_0 n \Delta n \qquad (2)$$

where $\omega_0$ is the optical mode frequency, n is the optical mode refractive index, and $\Delta n$ refers to the total refractive index change. The moving boundary effect, photo-elastic effect, and electro-optic effect are the three main mechanisms that cause refractive index changes ($\Delta n$) through propagating acoustic waves. Given the unsuspended structure of our AO devices, the contribution from the moving boundary effect is negligible. The contribution from electro-optic effect is also extremely weak, and it attributes to following factors. First, $SiO_2$ and polymer lack electro-optic coefficients. Second, although LN exhibits exceptional electro-optic coefficients along the z-axis of the crystal, our PCNBC is oriented along the z direction of the LN crystal, meaning the distribution of the TE mode in the waveguide aligns with the x direction of the LN crystal. More importantly, the electro-optic effect is an indirect contribution, highly dependent on the energy of the propagating acoustic waves. Combining the above-mentioned analyses, only the direct contribution to $\Delta n$ from the photo-elastic effect is significant in our devices. The photo-elastic effect relates $\Delta n$ to the photo-elastic coefficient and strain field. Therefore, $\Delta n$ can be written as[1,2],

$$\Delta n = \frac{n^3}{2} \cdot \frac{\iint (p_{11}S_1 + p_{12}S_2 + p_{14}S_4)|E_x|^2 dr}{\iint \varepsilon_r(r)|E_x|^2 dr} \qquad (3)$$

where $p_{ij}$ is the photo-elastic coefficient tensor and $S_j$ is the strain field sensor. When i and j take factors of 1, 2 and 4, they correspond to the x-direction, y-direction and shear yz-direction in the LN crystal, respectively. $|E_x|$ refers to the electric field component of the TE-polarized optical mode within the waveguide. Similarly, the contributions from the polymer and $SiO_2$ layers are considered. The photo-elastic coefficients ($p_{11}$, $p_{12}$, $p_{14}$) for the polymer, LN, and $SiO_2$ are (0.3, 0.3, 0)[37], (0.036, 0.135, 0.08)[27], and (0.121, 0.27, 0)[38], respectively. The distributions of optical electric field ($|E_x|$) and acoustic strain field ($S_{yy}$), calculated by the FEM, are shown in Figure 2a and Figure 2b, respectively. The polymer strip waveguide is 2 μm, and the period of the IDT is 4 μm. In this configuration, the refractive index n for the TE mode is calculated to be 1.83. Meanwhile, the designed IDT can successfully excite multiple acoustic modes. In Figure 2b, (i) and (ii) display only the dominant $S_{yy}$ strain components in the acoustic modes with the lowest (0.65 GHz) and highest (1.42 GHz) frequencies, respectively. Compared to 0.65 GHz, the strain field is more confined in the LN layer at 1.42 GHz. To illustrate the advantage of the PCNBC, the acoustic mode corresponding to 0.65 GHz is selected, and the calculated Δn is displayed in Table 1.

It is worth noting that $g_0$ is not associated with the performance of the optical and mechanical components in the AO modulator. To better understand the phonon-photon conversion, the photon number conversion efficiency η is introduced. At the acoustic mode resonance frequency $\Omega_m$, η can be expressed as[36,39],

$$\eta = 4C_0 n_c \eta_o \eta_m \qquad (4)$$

where $C_0 = 4g_0^2/\gamma\kappa$ is the single photon cooperativity, $n_c = 4\kappa_{ex}P_{opt}/[(4\Omega_m^2+\kappa^2)\hbar\omega_0]$ is the intracavity photon number of the blue-detuned pump light. $\eta_o = \kappa_{ex}/\kappa$ ($\eta_m = \gamma_{ex}/\gamma$) represents the external coupling efficiency of the optical (acoustic) modes, where $\kappa_{ex}$ ($\gamma_{ex}$) is the external coupling rate, and κ (γ) is the total optical (acoustic) mode linewidth. In addition, $P_{opt}$, $\Omega_m$, and ℏ correspond to the optical power that is sent into the cavity, the acoustic mode resonance frequency, and the reduced Planck constant, respectively.

To evaluate the cavity enhancement effect on the efficiency of AO modulation, we extract key parameters from numerical calculations and references to calculate η. Employing the parameter values summarized in Table 1, we estimate the photon number conversion efficiency η at various IDT aperture $L_a$, as depicted by solid lines in Figure 2c. This analysis assumes all the acoustic performance metrics remain constant. To further verify the key role of confinement factor Γ in AO modulation, we also examine the influence of a range of optical Q factors on η. As depicted by colored band in Figure 2c, the span between the upper and lower edges of each band represents the Q-induced variations in η. At the initial IDT aperture, due to the special hollow-field pattern, it can be observed that the second-order mode of PCNBC has almost no interaction with acoustic wave, thereby exhibiting a very low conversion efficiency, smaller than that of MRR. However, when the aperture exceeds 20 μm, there is a notable increase in Γ (see Figure 1c), leading to an enhanced η. In contrast, the fundamental mode demonstrates substantially high conversion efficiency at the initial stage with small IDT aperture, this result benefits from the compact Gaussian field distribution of the fundamental mode, leading to an efficient AO modulation. As depicted in the inset of Figure 2c, $\Delta\eta_{1st\text{-}MRR}$ and $\Delta\eta_{2nd\text{-}MRR}$ donate the enhancement in η for the fundamental and second-order modes of PCNBC relative to MRR, respectively. Note that $\Delta\eta_{1st\text{-}MRR}$ experiences a large value from 20.7 dB to 19.8 dB as $L_a$ expands from 10 μm to 40 μm, but it has to suffer a low η representing a limited photon conversion efficiency in this aperture range. In comparison, when the growth in Γ for the fundamental mode of PCNBC tends to plateau, whereas it continues linearly for MRR, causing $\Delta\eta_{1st\text{-}MRR}$ to continuously drop down below 15.8 dB. Meanwhile, $\Delta\eta_{2nd\text{-}MRR}$ remains relatively stable around 9 dB, due to a comparable Q and transmission of second-order mode of PCNBC and MRR. Our investigation into the cavity-enhanced interaction between the fundamental mode of PCNBC and MRR identifies an optimal aperture range from 40 μm to 80 μm that nearly maximizes η while maintaining a high $\Delta\eta_{1st\text{-}MRR}$, as highlighted in Figure 2c. This balance emphasizes the complicated relationship between acoustic transducer dimension and AO modulation efficiency, opening the door for cavity-enhanced devices.

**Table 1. Parameter set for evaluation of the photon number conversion efficiency η**

| Parameters | | MRR | PCNBC 2nd mode | PCNBC 1st mode | Units |
|---|---|---|---|---|---|
| IDT[40] | $\Omega_m$ | 0.65 | 0.65 | 0.65 | GHz/2π |
| | $\gamma$ | 6.7 | 6.7 | 6.7 | MHz/2π |
| | $\gamma_{ex}/\gamma$ | 0.015 | 0.015 | 0.015 | / |
| Optical cavity[19] | Transmission | 0.99 | 0.60 | 0.15 | / |
| | Q | 2 | 2 | 5 | ×10^4 |
| | $\kappa$ | 9.67 | 9.67 | 3.87 | GHz/2π |
| | $\kappa_{ex}/\kappa$ | 0.5 | 0.39 | 0.19 | / |
| | $P_{opt}$ | 0.2 | 0.2 | 0.2 | mW |
| | $\omega_0$ | 193.54 | 193.17 | 193.54 | THz/2π |
| AO interaction | $\Delta n$ | 2.2 | 2.2 | 2.2 | ×10^{-14} |

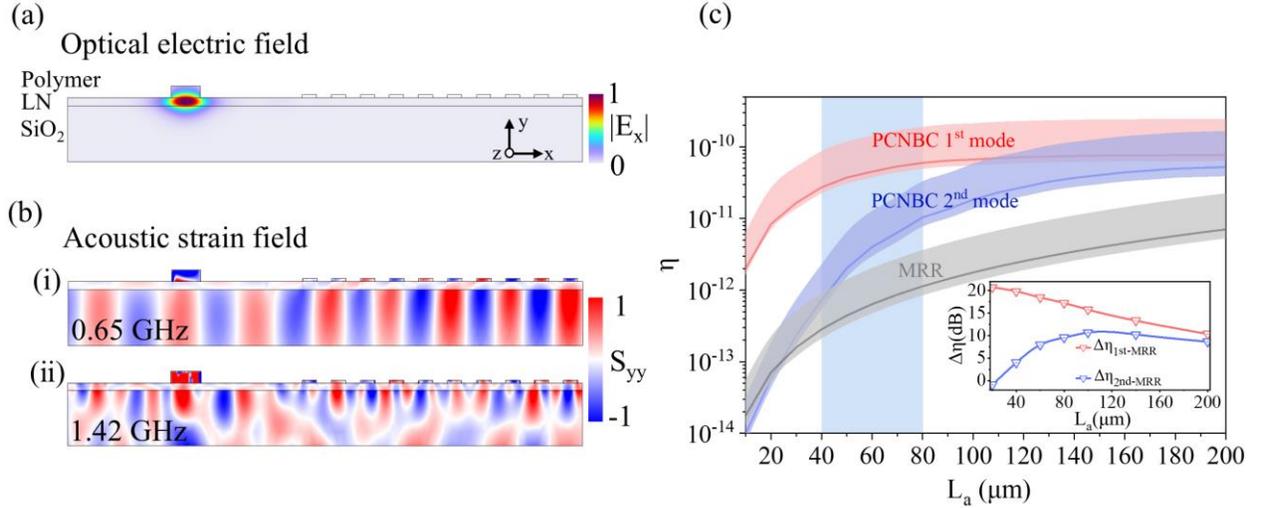

**Figure 2.** (a) A portion of the cross-sectional schematic of the AO modulator for the polymer-loaded LN integrated platform shows the TE polarized optical mode field distribution at the wavelength of 1550 nm, as depicted by the color profile. The polymer strip waveguide has a width of 2 μm and a thickness of 400 nm. (b) (i) and (ii) depict the acoustic strain field distributions ($S_{yy}$) caused by acoustic modes at 0.65 GHz and 1.42 GHz, respectively. The period and thickness of the IDT are 4 μm and 100 nm, respectively. Extending along the x direction places 10 IDT finger pairs. The strain field is normalized in both frequencies. (c) Calculated η for AO PCNBC and MRR as a function of $L_a$. The Q factor ranges are $10^4$ to $8\times10^4$ for the fundamental mode and $10^4$ to $3\times10^4$ for both the second-order mode and MRR. The solid lines within each color band correspond to the Q factors in Table 1.

The above analysis inspires an investigation into cavity enhancement in AO modulators, aiming to provide a universal enhancement solution for various AO devices. We developed our devices on a polymer-loaded LN integrated platform, specifically designing a PCNBC that operates in TE-polarized mode. This design is critical as it ensures the propagation of standard bound modes without radiation losses for any width within their supported range[41], resulting in optimal mode overlapping with RF channel design in IDT part. In addition to above-mentioned PCNBC, a comparative reference device is conducted with a MRR featuring a 200 μm bending radius and a 60 μm straight waveguide segment. Both the PCNBC and MRR are incorporated with identical IDT with a period of 4 μm and aperture $L_a$ of 60 μm. This design is aiming at overlapping with the majority of photons confined in the cavity, particularly serving for the fundamental mode of PCNBC.

## ■ METHODS

The devices were fabricated on a customized y-cut LNOI wafer purchased from NanoLN Corporation. The process began with defining the IDTs using e-beam lithography, followed by e-beam evaporation to deposit a 5 nm thick titanium (Ti) adhesion layer and a 95 nm gold (Au) layer. The resist residue was removed by immersing the chip in the organic solvent N-Methyl-2-pyrrolidone. A 400 nm thick resist (e-beam positive photoresist, ARP-6200) was spin-coated onto the wafer as the polymer loading material. A second exposure was used to pattern all the optical components, including grating couplers, transmission waveguides, and optical cavities.

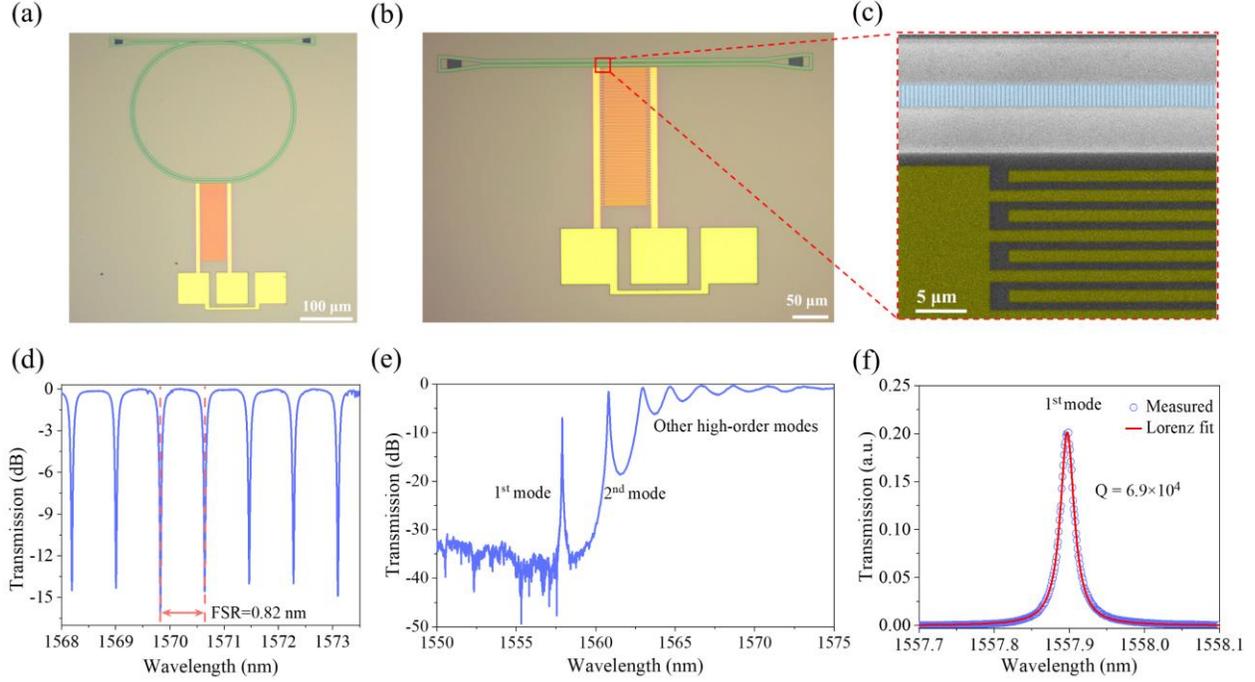

**Figure 3.** (a) Optical microscope image of the MRR-based AO modulator. (b) Optical microscope image of the PCNBC-based AO modulator. The IDT has the same parameters as the AO MRR. (c) False-color SEM image of the red rectangular region in (b). (d) Measured optical transmission spectrum of the MRR from 1568.0 nm to 1573.5 nm. (e) Measured optical transmission spectrum of the PCNBC. The fundamental mode and second-order mode locate at 1557.9 nm and 1560.8 nm, respectively. (f) Zoom-in of the fundamental mode and fitted by Lorentzian line shape, showing the Q factor of $6.9\times10^4$.

Figure 3a and 3b display the microscope images of the fabricated MRR- and PCNBC-based AO modulators, respectively. Light is coupled into and out of the optical cavities through a pair of grating couplers with a period of 990 nm and a filling factor of 0.5. In Figure 3c, a magnified false-color scanning electron microscope (SEM) image of the PCNBC and IDT is shown, corresponding to the red box region in Figure 3b. It shows the fabricated IDT electrodes with a period of 4 μm, which is twice the waveguide width of PCNBC. Meanwhile, to avoid optical loss due to metal absorption, the distance d between IDT and the PCNBC is maintained at 6.5 μm. Figure 3d and 3e present the measured optical transmission spectra for two types cavities over a broad wavelength range. In Figure 3d, the MRR has a measured free spectral range (FSR) of 0.82 nm. Based on the measurement results, the group index $n_g$ is related to the measured FSR by $n_g = \lambda^2/(FSR \cdot L_c)$, in which $\lambda$ is the wavelength[42]. The solved value of $n_g$ is 2.18, which agrees well with the predicted value of 2.18 for TE mode calculated by the FEM model. Moreover, the MRR has a Q factor of $1.8 \times 10^4$ and an extinction ratio of 16 dB at the resonance peaks of 1569.8 nm. Figure 3e shows that the PCNBC exhibits several optical resonances, where the fundamental mode and the second-order mode correspond to the resonance peaks at 1557.9 nm and 1560.8 nm, respectively. The detailed transmission spectrum characterization of the fundamental mode (Figure 3f) exhibits a Q factor of $6.9 \times 10^4$, corresponding to a transmission of 0.2. For the second-order mode, it exhibits a lower Q factor of $1.6 \times 10^4$ but a larger transmission of 0.7. Two possible reasons for the decreased fundamental mode transmission are the cavity losses (intrinsic and coupling loss), and the phenomenon of SHG resulting from the high energy density confined in nonlinear medium[28].

## ■ RESULTS AND DISCUSSIONS

The characterization of our AO devices is performed using the experimental setup depicted in Figure 3a. Light emitted from a tunable semiconductor laser (TSL, Santec/TSL550) passes through a polarization controller (PC) before being coupled into the optical cavities via a grating coupler. Meanwhile, the RF signal from Port 1 of a calibrated VNA (Keysight/E5071C) is transmitted to the IDT through a microwave probe, as illustrated in the inset optical image. The modulated optical signal is coupled out from the cavities by another grating coupler and then amplified by an erbium-doped fiber amplifier (EDFA) to ensure the detection by the high-sensitivity PD (Finisar/XPDV2120RA). Before sending the output signal to PD, a band-pass filter (BPF) between the EDFA and PD is used to remove noise and perturbation pulses induced by the EDFA. Finally, the output electrical signal from PD is fed back to Port 2. To prevent the perturbation of ambient temperature, the chip is placed on a high-accuracy thermoelectric cooler with a stable temperature of 22°C.

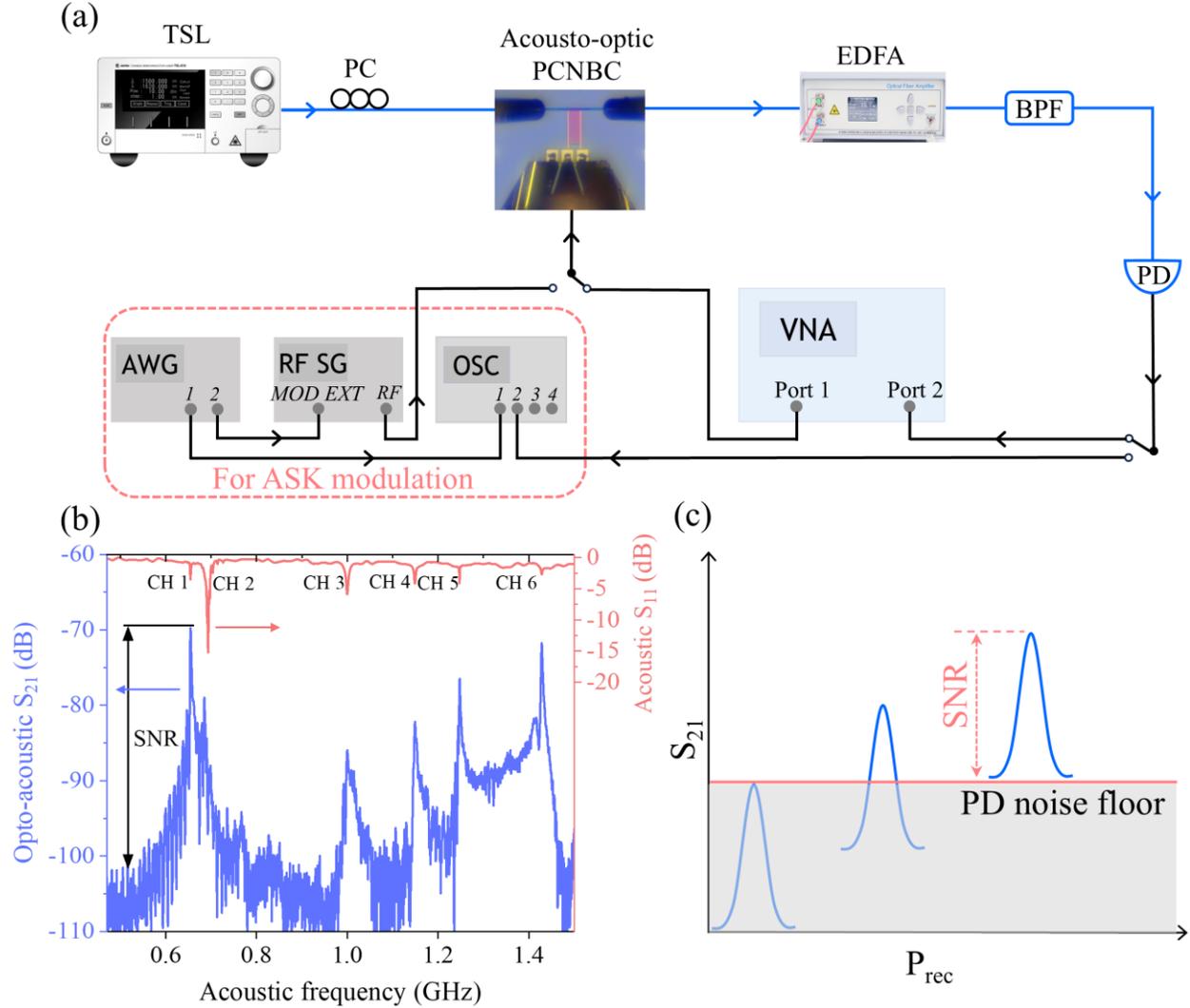

**Figure 4.** (a) Schematic illustration of the experimental testing setup, the red dashed box is used to demonstrate the AO digital modulation. TLS, tunable laser source; PC, polarization controller; EDFA, erbium-doped fiber amplifier; BPF, bandpass filter; PD, photodetector; VNA, vector network analyzer; AWG, arbitrary waveform generator; RF SG, radio frequency signal generator; OSC, oscilloscope. (b) Measured $S_{11}$ and the opto-acoustic $S_{21}$ spectra for the fundamental mode of the PCNBC. The PD received power $I_{rec} = 0.1$ mW after being amplified

by EDFA, and the output power of the VNA is $P_{RF}$ = 5 dBm. (c) Schematic illustration of the signal reception at different optical power levels.

During the measurements, the optical power launched into the PCNBC is maintained at 87 μW to prevent the excitation of nonlinear optical effects. Simultaneously, the output power ($P_{RF}$) of VNA Port 1 is set to 5 dBm to provide sufficient energy for the acoustic waves excited by IDT to deform the waveguide. The received optical power ($P_{rec}$) of PD is set to 0.1 mW. Figure 4b plots the measured $S_{11}$ and $S_{21}$ spectra with optical channel at the fundamental mode. The $S_{11}$ spectrum has six prominent dips in the range between 0.4 and 1.5 GHz that reveal acoustic modes excited by the IDT. The number of the acoustic mode corresponds to the number of modulation channels (CH), and the frequencies of CH 1 and CH 2 agree well with the FEM results. On the one hand, the acoustic performance parameters can be extracted from the $S_{11}$ spectrum. For example, at 0.65 GHz, the microwave-to-acoustic coupling efficiency 1-$|S_{11}|^2$ is 52%, with a measured acoustic Q factor of 425. To characterize the AO interaction, the pump laser is blue-detuned from the fundamental mode resonance wavelength. The $S_{21}$ spectrum exhibits six peaks that match the acoustic resonance frequencies in the $S_{11}$ spectrum, indicating that the optical resonance of the fundamental mode is effectively modulated by these acoustic modes through the photo-elastic effect. Note that the $S_{21}$ spectrum reveals the AO response in the frequency domain. In addition, η reveals the intra-cavity photon number conversion efficiency from input microwave photons to output sideband photons coupled out of the cavity, and it highlights the inherent performance of the AO device. Here, we relate η to the measured $S_{21}$ using the PD, and $S_{21}$ can be described as[16],

$$S_{21} = \frac{P_{PD}}{P_{RF}} = \eta \cdot \frac{2R_{PD}^2 R_{load} \omega_0 P_{opt}}{\Omega_m} \tag{5}$$

where $P_{PD}$ and $R_{PD}$ are the output microwave power and responsivity (A/W) of the PD, respectively. $R_{load}$ represents the impedance of the microwave transmission line. Note that equation (4) represents the peak η when the driven RF frequency equals the acoustic mode resonance frequency. This means that the measured SNR of the $S_{21}$ spectrum directly reveals the AO modulation efficiency. A higher SNR is of significant interest to sustain the signal quality, especially in integrated microwave photonics (MWP)[43]. Nevertheless, even with the excellent optical and acoustic performance of the AO device, there is still a challenge in directly capturing the true SNR of the $S_{21}$. As shown in Figure 4c, the noise floor indicates that the PD has a minimum measurable power. The optical signal can be perfectly converted into an electric signal only if its minimum power exceeds the minimum measurable power of the PD.

To quantify the AO modulation efficiency, we fix the drive frequency of IDTs at approximately 0.65 GHz for testing. This specific frequency is selected for its higher SNR compared to other frequencies within the $S_{21}$ spectrum, as shown in Figure 4b. Subsequently, we extract the SNR values from the measured $S_{21}$ at $P_{RF}$ from -50 dBm to 5dBm and $P_{rec}$ from 0.1 mW to 1.5 mW, as depicted in Figure 5a. The variations in color intensity within these figures highlight differences in modulation efficiency. As anticipated, the fundamental mode and second-order mode of PCNBC exhibits both cavity enhancement effect in SNR and $P_{RF}$ compared with MRR. The fundamental mode of PCNBC has the highest saturated SNR value of 38 dB when $P_{RF}$ increases to above 0 dBm, while the second-order mode of PCNBC and the MRR have values of 33 dB and 29 dB, respectively. Furthermore, we depict the SNR data of those three mechanisms at the same optical level of $P_{rec}$ = 1.2 mW for comparison (Figure 5b). The threshold RF power $P_{th}$, which donates minimum RF power required to enable the modulation, is another critical parameter in the operation of AO modulator. According to the Equation (5), $P_{th} = P_{noise}/S_{21}$, where $P_{noise}$ represents the background noise power of the conversion system. Consequently, $P_{th}$ is found to be inversely proportional to η. By fitting the measured data, the fundamental mode of PCNBC is capable of operating at RF power levels as low as -50 dBm (Figure 5b), which is a very attractive value for low power applications. For the enhancement effect investigation, $\Delta P_{th}$ between the fundamental mode of PCNBC and the MRR is further calculated versus $P_{rec}$, showing stable value around -22.5 dB, which matches well with the theoretically calculated $\Delta\eta_{1st\text{-}MRR}$ of 23.2 dB obtained by updating the measured parameters to Equation (4). Similarly, the extracted $\Delta P_{th}$ between the fundamental mode and the second-order mode, as well as between the second-order mode and MRR, are about 15.5 dB and 6.6 dB, which are in close agreement with the calculated $\Delta\eta_{1st\text{-}MRR}$ of 15.6 dB and 7.7 dB, respectively. The slight deviation of the extracted $\Delta P_{th}$ from the calculated factors can be attributed to differences in the Q factors of PCNBC between measurement and simulation, which induce a small variation in Γ.

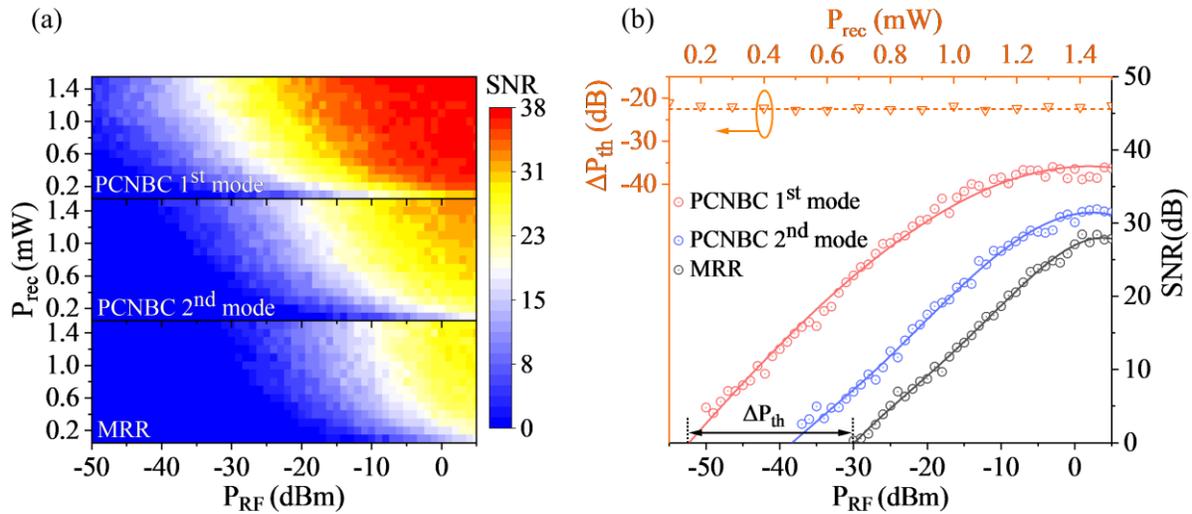

**Figure 5.** (a) The extracted SNR as functions of $P_{RF}$ and $P_{rec}$ for the fundamental mode, the second-order mode of PCNBC and MRR. (b) Dependence of the SNR on the $P_{RF}$ when $P_{rec}$ is fixed at 1.2 mW. Meanwhile, we plot the $\Delta P_{th}$ between the fundamental mode and MRR as a function of $P_{rec}$. The extracted modulation $\Delta P_{th}$ is about -22.5 dB.

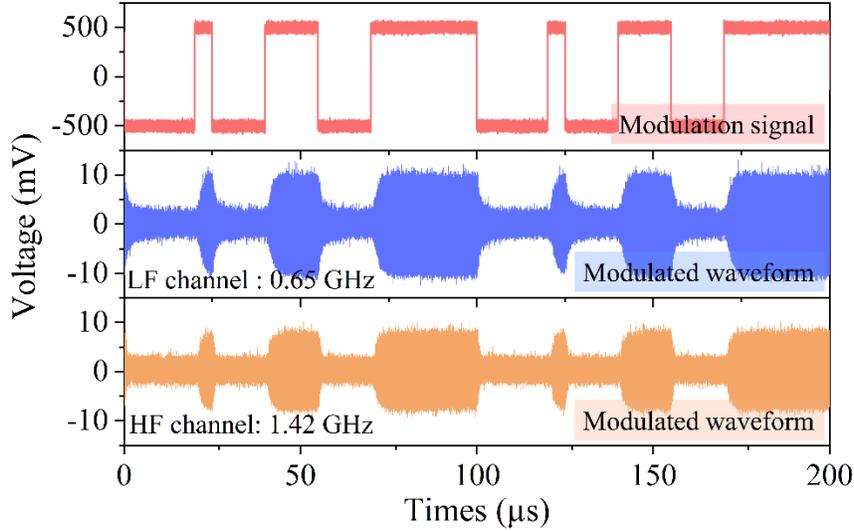

**Figure 6** Demonstration of the ASK modulation based on the AO PCNBC that operates in the fundamental mode at various carrier frequencies: a low frequency (LF) channel at 0.65 GHz and a high frequency (HF) channel at 1.42 GHz. The modulation signal is generated by the AWG.

Current MWP systems on chip urgently require a monolithic device that integrates both filtering and modulation functions to replace the complex heterogeneous integration of electrical filters and EO modulators.[44] In wireless telecommunications, the launched or received RF signal by the antenna is predominantly based on a modulated digital signal, such as amplitude shift keying (ASK), phase shift keying (PSK), and quadrature amplitude modulation (QAM), owing to its low susceptibility to noise and waveform distortion.[45] To demonstrate the fundamental conversion capabilities of our cavity-enhanced AO modulator, we employed the experimental setup with source equipment highlighted in red dashed box in Figure 3a. In this setup, the IDT is driven by a digital RF signal generated through ASK modulation, facilitated by the use of arbitrary waveform generator (AWG, Tektronix/AWG710) and RF signal generator (SG, Rohde & Schwarz/SMB100A). Due to the high SNR and energy efficiency of the PCNBC-based AO modulator that operates in the fundamental mode, it is particularly suitable for the MWP. As shown in Figure 6, time-domain traces of the output modulated optical signals at carrier frequencies of 0.65 GHz and 1.42 GHz closely correspond to the reference digital modulation signal, effectively validating the capability of our devices to convert narrow-band RF signals to optical signals through nanophotonic structure on LNOI substrate. This verification holds promising potential for applications in MWP systems and optical modules that require efficient and compact EO conversion.

■ **CONCLUSIONS**

In summary, we present the first experimental demonstration of the cavity enhancement effect in AO modulators developed on an integrated lithium niobate platform. The fabricated PCNBC-based AO modulator exhibits a significantly enhanced SNR of 38 dB and operates with a threshold RF power below -50 dBm, which is two orders of magnitude lower than MRR-based devices. In addition, robust ASK microwave to optical modulation at channels of 0.65 GHz and 1.42 GHz are also vitrificated. These achievements represent a significant progress in the development of highly efficient, chip-scale AO functional devices, enabling a universal technology for high performance realization. The unique capability of PCNBCs to provide almost ideal light-sound confinement factor makes them as a compelling photonic platform, establishing it as a promising candidate for on-chip integrated microwave photonics, optical transceivers, and computing applications.

**Author Contributions**

The manuscript was written through contributions of all authors. All authors have given approval to the final version of the manuscript.

**Notes**

The authors declare no competing financial interest.

**ACKNOWLEDGMENT**

The authors acknowledge support from the National Key R&D Program of China (No. 2022ZD0119002), the National Natural Science Foundation of China (Grant No. 62025402, 62090033, 92364204, 92264202 and 62293522) and Major Program of Zhejiang Natural Science Foundation (Grant No. LDT23F04024F04)

**REFERENCES**

(1) Cai, L. T.; Mahmoud, A.; Khan, M.; Mahmoud, M.; Mukherjee, T.; Bain, J.; Piazza, G. Acousto-optical modulation of thin film lithium niobate waveguide devices. *Photonics Res.* **2019**, 7 (9), 1003-1013.

(2) Tadesse, S. A.; Li, M. Sub-optical wavelength acoustic wave modulation of integrated photonic resonators at microwave frequencies. *Nat. Commun.* **2014**, 5 (1), 5402.

(3) Wan, L.; Yang, Z.; Zhou, W.; Wen, M.; Feng, T.; Zeng, S.; Liu, D.; Li, H.; Pan, J.; Zhu, N.; et al. Highly efficient acousto-optic modulation using nonsuspended thin-film lithium niobate-chalcogenide hybrid waveguides. *Light-Sci. Appl.* **2022**, 11 (1), 145.

(4) Sarabalis, C. J.; Van Laer, R.; Patel, R. N.; Dahmani, Y. D.; Jiang, W.; Mayor, F. M.; Safavi-Naeini, A. H. Acousto-optic modulation of a wavelength-scale waveguide. *Optica* **2021**, 8 (4), 477-483.

(5) Shao, L. B.; Sinclair, N.; Leatham, J.; Hu, Y. W.; Yu, M. J.; Turpin, T.; Crowe, D.; Loncar, M. Integrated microwave acousto-optic frequency shifter on thin-film lithium niobate. *Opt. Express* **2020**, 28 (16), 23728-23738.

(6) Yu, Z.; Sun, X. Gigahertz Acousto-Optic Modulation and Frequency Shifting on Etchless Lithium Niobate Integrated Platform. *ACS Photonics* **2021**, 8 (3), 798-803.

(7) Kittlaus, E. A.; Jones, W. M.; Rakich, P. T.; Otterstrom, N. T.; Muller, R. E.; Rais-Zadeh, M.

Electrically driven acousto-optics and broadband non-reciprocity in silicon photonics. *Nat. Photonics* **2020**, 15 (1), 43-52.

(8) Sohn, D. B.; Bahl, G. Direction reconfigurable nonreciprocal acousto-optic modulator on chip. *APL Photonics* **2019**, 4 (12), 126103.

(9) Huang, C.; Shi, H.; Yu, L.; Wang, K.; Cheng, M.; Huang, Q.; Jiao, W.; Sun, J. Acousto-Optic Modulation in Silicon Waveguides Based on Piezoelectric Aluminum Scandium Nitride Film. *Adv. Opt. Mater.* **2022**, 10 (6), 2102334.

(10) Zhao, H.; Li, B.; Li, H.; Li, M. Enabling scalable optical computing in synthetic frequency dimension using integrated cavity acousto-optics. *Nat. Commun.* **2022**, 13 (1), 5426.

(11) Xu, X.-B.; Wang, W.-T.; Sun, L.-Y.; Zou, C.-L. Hybrid superconducting photonic-phononic chip for quantum information processing. *Chip* **2022**, 1 (3), 100016.

(12) Tadesse, S. A.; Li, H.; Liu, Q.; Li, M. Acousto-optic modulation of a photonic crystal nanocavity with Lamb waves in microwave K band. *Appl. Phys. Lett.* **2015**, 107 (20), 201113.

(13) Ghosh, S.; Piazza, G. Laterally vibrating resonator based elasto-optic modulation in aluminum nitride. *APL Photonics* **2016**, 1 (3), 036101.

(14) Yang, Z.; Wen, M.; Wan, L.; Feng, T.; Zhou, W.; Liu, D.; Zeng, S.; Yang, S.; Li, Z. Efficient acousto-optic modulation using a microring resonator on a thin-film lithium niobate-chalcogenide hybrid platform. *Opt. Lett.* **2022**, 47 (15), 3808-3811.

(15) Li, H.; Tadesse, S. A.; Liu, Q.; Li, M. Nanophotonic cavity optomechanics with propagating acoustic waves at frequencies up to 12 GHz. *Optica* **2015**, 2 (9), 826-831.

(16) Shao, L.; Yu, M.; Maity, S.; Sinclair, N.; Zheng, L.; Chia, C.; Shams-Ansari, A.; Wang, C.; Zhang, M.; Lai, K.; et al. Microwave-to-optical conversion using lithium niobate thin-film acoustic resonators. *Optica* **2019**, 6 (12), 1498-1505.

(17) Lu, R.; Manzaneque, T.; Yang, Y.; Li, M. H.; Gong, S. Gigahertz Low-Loss and Wideband S0 Mode Lithium Niobate Acoustic Delay Lines. *IEEE Trans Ultrason Ferroelectr Freq Control* **2019**, 66, 1373-1386.

(18) Li, M.; Ling, J.; He, Y.; Javid, U. A.; Xue, S.; Lin, Q. Lithium niobate photonic-crystal electro-optic modulator. *Nat. Commun.* **2020**, 11 (1), 4123.

(19) Zhang, J.; Pan, B.; Liu, W.; Dai, D.; Shi, Y. Ultra-compact electro-optic modulator based on etchless lithium niobate photonic crystal nanobeam cavity. *Opt. Express* **2022**, 30 (12), 20839-20846.

(20) Dmitriev, V.; Kawakatsu, M. N.; Portela, G. Compact optical switch based on 2D photonic crystal and magneto-optical cavity. *Opt. Lett.* **2013**, 38 (7), 1016-1018.

(21) Shakoor, A.; Nozaki, K.; Kuramochi, E.; Nishiguchi, K.; Shinya, A.; Notomi, M. Compact 1D-silicon photonic crystal electro-optic modulator operating with ultra-low switching voltage and energy. *Opt. Express* **2014**, 22 (23), 28623-28634.

(22) Zhou, H.; Qiu, C.; Jiang, X.; Zhu, Q.; He, Y.; Zhang, Y.; Su, Y.; Soref, R. Compact, submilliwatt, 2 × 2 silicon thermo-optic switch based on photonic crystal nanobeam cavities. *Photonics Res.* **2017**, 5 (2), 108-112.

(23) Zhang, Y.; He, Y.; Zhu, Q.; Guo, X.; Qiu, C.; Su, Y.; Soref, R. Single-resonance silicon nanobeam filter with an ultra-high thermo-optic tuning efficiency over a wide continuous tuning range. *Opt. Lett.* **2018**, 43 (18), 4518-4521.

(24) Quan, Q.; Burgess, I. B.; Tang, S. K.; Floyd, D. L.; Loncar, M. High-Q, low index-contrast polymeric photonic crystal nanobeam cavities. *Opt. Express* **2011**, 19 (22), 22191-22197.

(25) Yao, D.; Jiang, Z.; Zhang, Y.; Xie, H.; Wang, T.; Wang, J.; Gan, X.; Han, G.; Liu, Y.; Hao, Y. Ultrahigh thermal-efficient all-optical silicon photonic crystal nanobeam cavity modulator with TPA-induced thermo-optic effect. *Opt. Lett.* **2023**, 48 (9), 2325-2328.

(26) Quan, Q. M.; Deotare, P. B.; Loncar, M. Photonic crystal nanobeam cavity strongly coupled to the feeding waveguide. *Appl. Phys. Lett.* **2010**, 96 (20), 203102.

(27) Weis, R. S.; Gaylord, T. K. Lithium niobate: Summary of physical properties and crystal structure. *Applied Physics A Solids and Surfaces* **1985**, 37, 191-203.

(28) Zhu, D.; Shao, L. B.; Yu, M. J.; Cheng, R.; Desiatov, B.; Xin, C. J.; Hu, Y. W.; Holzgrafe, J.; Ghosh, S.; Shams-Ansari, A.; et al. Integrated photonics on thin-film lithium niobate. *Adv. Opt. Photonics* **2021**, 13 (2), 242-352.

(29) Yu, Z. J.; Xi, X.; Ma, J. W.; Tsang, H. K.; Zou, C. L.; Sun, X. K. Photonic integrated circuits with bound states in the continuum. *Optica* **2019**, 6 (10), 1342-1348.

(30) Ye, F.; Yu, Y.; Xi, X.; Sun, X. Second-Harmonic Generation in Etchless Lithium Niobate Nanophotonic Waveguides with Bound States in the Continuum. *Laser Photon. Rev.* **2022**, 16 (3), 2100429.

(31) Yu, Z.; Tong, Y.; Tsang, H. K.; Sun, X. High-dimensional communication on etchless lithium niobate platform with photonic bound states in the continuum. *Nat. Commun.* **2020**, 11 (1), 2602.

(32) Yu, Y.; Yu, Z. J.; Zhang, Z. Y.; Tsang, H. K.; Sun, X. K. Wavelength-Division Multiplexing on an Etchless Lithium Niobate Integrated Platform. *ACS Photonics* **2022**, 9 (10), 3253-3259.

(33) Yu, Z.; Sun, X. Acousto-optic modulation of photonic bound state in the continuum. *Light-Sci. Appl.* **2020**, 9 (1), 1.

(34) Yu, Y.; Wang, L.; Sun, X. Demonstration of on-chip gigahertz acousto-optic modulation at near-visible wavelengths. *Nanophotonics* **2021**, 10 (17), 4323-4329.

(35) Jiang, Z.; Fang, C.; Ran, X.; Gao, Y.; Wang, R.; Wang, J.; Yao, D.; Gan, X.; Liu, Y.; Hao, Y.; et al. Ultra-high-Q photonic crystal nanobeam cavity for etchless lithium niobate on insulator

(LNOI) platform. *Opto-Electron. Adv.* **2025**, 0, 240114-240114.

(36) Aspelmeyer, M.; Kippenberg, T. J.; Marquardt, F. Cavity optomechanics. *Rev. Mod. Phys.* **2014**, 86 (4), 1391-1452.

(37) Jena, S.; Tokas, R. B.; Thakur, S.; Udupa, D. V. Tunable mirrors and filters in 1D photonic crystals containing polymers. *Physica E: Low-dimensional Systems and Nanostructures* **2019**, 114, 113627.

(38) Hah, D.; Yoon, E.; Hong, S. An optomechanical pressure sensor using multimode interference couplers with polymer waveguides on a thin p+-Si membrane. *Sens. Actuator A-Phys.* **2000**, 79 (3), 204-210.

(39) Jiang, W.; Sarabalis, C. J.; Dahmani, Y. D.; Patel, R. N.; Mayor, F. M.; Mckenna, T. P.; Van Laer, R.; Safavi-Naeini, A. H. Efficient bidirectional piezo-optomechanical transduction between microwave and optical frequency. *Nat. Commun.* **2020**, 11 (1), 1166.

(40) Marinković, I.; Drimmer, M.; Hensen, B.; Gröblacher, S. Hybrid Integration of Silicon Photonic Devices on Lithium Niobate for Optomechanical Wavelength Conversion. *Nano Lett.* **2021**, 21 (1), 529-535.

(41) Ctyroky, J.; Petracek, J.; Kuzmiak, V.; Richter, I. Bound modes in the continuum in integrated photonic LiNbO$_3$ waveguides: are they always beneficial? *Opt. Express* **2023**, 31 (1), 44-55.

(42) Rabiei, P.; Steier, W. H.; Zhang, C.; Dalton, L. R. Polymer micro-ring filters and modulators. *J. Lightwave Technol.* **2002**, 20 (11), 1968-1975.

(43) Liu, Y.; Choudhary, A.; Marpaung, D.; Eggleton, B. J. Integrated microwave photonic filters. *Adv. Opt. Photonics* **2020**, 12 (2), 485-555.

(44) Marpaung, D.; Yao, J.; Capmany, J. Integrated microwave photonics. *Nat. Photonics* **2019**, 13 (2), 80-90.

(45) Kadir, E. A.; Shubair, R.; Abdul Rahim, S. K.; Himdi, M.; Kamarudin, M. R.; Rosa, S. L. B5G and 6G: Next Generation Wireless Communications Technologies, Demand and Challenges. *Journal* **2021**, 1-6.